\documentclass[12pt,preprint]{aastex}

\newcommand{\lhs} {LHS 3250}
\newcommand{\sdss} {SDSS 1337+00}
\newcommand{\logg} {\log g}

\newcommand{\Te} {T_{\rm eff}}

\newcommand{\mv} {$M_V$}
\newcommand{\htwo} {H$_2$}
\newcommand{\msun} {$M_\odot$}

\newcommand\gta{\lower 0.5ex\hbox{$\buildrel > \over \sim\ $}} %greater than about
\newcommand\lta{\lower 0.5ex\hbox{$\buildrel < \over \sim\ $}} %less than about
\newcommand{\nhe} {N({\rm He})/N({\rm H})}
\newcommand{\nh} {N({\rm H})/N({\rm He})}
\shortauthors{Bergeron \& Leggett}
\shorttitle{Analysis of Cool White Dwarfs}
\begin{document}

%\title{\version}

\title{Model Atmosphere Analysis of Two Very Cool White Dwarfs}

\author{P. Bergeron}
\affil{D\'epartement de Physique, Universit\'e de Montr\'eal, C.P.~6128, 
Succ.~Centre-Ville, 
Montr\'eal, Qu\'ebec, Canada, H3C 3J7.}
\email{bergeron@astro.umontreal.ca}
\and
\author{S.~K. Leggett}
\affil{UKIRT, Joint Astronomy Centre, 660 North A'ohoku Place, Hilo, HI 96720.}
\email{s.leggett@jach.hawaii.edu}

\begin{abstract}

A detailed analysis of the very cool white dwarfs \sdss\ and \lhs\ is
presented. Model atmosphere calculations with improved
collision-induced absorptions by molecular hydrogen indicate that a
pure hydrogen composition can be ruled out, and that the strong
infrared absorption observed in these cool stars is better explained
in terms of collisions of \htwo\ with neutral helium. It is shown that
even though the overall shape of the observed energy distributions can
be reproduced reasonably well with helium-rich models, the peak of the
energy distribution near 6000 \AA\ is always predicted too narrow.
The extreme helium-rich composition inferred for both objects is
discussed in the broader context of the extremely cool white dwarfs
reported in various surveys.

\end{abstract}


\section{Introduction}

The photometric and spectroscopic analyses of \citet{brl97,blr01} have
revealed that the coolest white dwarfs observed in the Galactic disk
have effective temperatures in excess of $\Te\sim4000$~K \citep[see,
e.g., Fig.~21 of][]{blr01}. The absence of any cooler white dwarfs can
be interpreted as the result of the finite age of the disk, which has
been estimated from a determination of the luminosity function at 8
$\pm$ 1.5 Gyr \citep{lrb98}. This value is also consistent with the
location of white dwarfs in a mass versus effective temperature
diagram in which theoretical isochrones are overplotted \citep[see
Figs.~24 and 25 of][]{blr01}. Since the Galactic halo is believed to
have formed many Gyr earlier than the disk, white dwarfs associated
with the halo could be considerably older and thus cooler than those
found in the disk. In addition, halo white dwarfs could be easily
identified by their peculiar kinematics. This was first recognized by
\citet{ldm89} who identified 6 white dwarfs in their luminosity
function sample that had tangential velocities consistent with a halo
population ($v_{\rm tan}\ \gta250$ km~s$^{-1}$). Several objects from
that small sample are now believed to be too young (they are too warm
and massive) to belong to the Galactic halo, however
\citep{fon01}.

More recently, white dwarfs with effective temperatures below 4000~K
have been identified in various surveys
\citep{hambly99,harris99,ibata00,harris01,opp01a,opp01b,ruiz01,scholz02,farihi02},
many of which are believed to be associated with the halo population.
Even though the model fluxes are successful at reproducing in detail
the optical and infrared broadband photometric observations of white
dwarfs above 4000~K \citep{brl97,blr01}, the observed energy
distributions of cooler white dwarfs are at odds with the predictions
of model atmosphere calculations. For instance, the analysis of \lhs\
by
\citet{harris99} has shown that the observed optical and infrared 
energy distribution could not be reproduced adequately in terms of a
pure hydrogen atmosphere, or a mixed hydrogen and helium
composition. Similar conclusions were reached by
\citet{opp01b}. For WD 0346+246 (also analyzed by Oppenheimer et al.),
\citet{ber01} was even forced to introduce an {\it ad hoc} source of
opacity to reproduce the observed photometry at $B$, $V$, and $R$.

\sdss\ is another extremely cool white dwarf discovered by \citet{harris01} 
in imaging data from the Sloan Digital Sky Survey. The optical and
near-infrared spectrum of this object resemble that of
\lhs\ \citep[see Fig.~3 of][]{harris01}. Photometric observations reported
by Harris et al.~covered only the optical $BVRI$ and a lower limit on
the infrared $J$ magnitude. In this paper, we report new infrared
photometric observations at $J$ and $H$ for \sdss, and provide a
thorough analysis of this object as well as of \lhs, its almost
identical twin. Both of these objects exhibit the strong infrared flux
deficiency that results from collision-induced absorptions by
molecular hydrogen. New calculations of this important source of
opacity by Borysow and collaborators has led us to take a fresh look
at the atmospheric properties of cool white dwarf atmospheres. In this
paper, we thus explore in detail the effects of effective temperature,
surface gravity, and atmospheric composition (hydrogen, helium, and
heavier elements) on the predicted fluxes, and compare these emergent
flux distributions with those of \sdss\ and
\lhs.

\section{Observations}

The optical $BVRI$ photometric observations for \lhs\ and \sdss\ are
taken from \citet{harris99} and \citet{harris01}, respectively.  The
infrared $JHK$ photometry for \lhs\ is also taken from
\citet{harris99}, while new $J$ and $H$ photometric observations were
secured for \sdss\ on the nights of 2001 January 9th and 31st (UT)
respectively, using the UFTI camera on the UK Infrared Telescope
(UKIRT) on Mauna Kea Hawaii. Total observation time was 72 minutes for
$J$ and 63 minutes for $H$, made up in both cases of 60 second
exposures in a repeated 9-position dither pattern. UKIRT Faint
Standards were used to calibrate the photometry.  The derived
magnitudes were $J=20.38\pm0.09$ and $H=20.71\pm0.15$, on the Mauna
Kea photometric system. 

The energy distributions of \sdss\ and \lhs\ have been converted into
broadband fluxes following the prescription of \citet{brl97}. These
are shown in the upper two panels of Figure \ref{fg:f1}, together with
the spectroscopic observations discussed in \citet{harris01}, and
kindly made available to us by H. Harris. It is clear from this
comparison, at least in the case of
\sdss, that a detailed comparison of the observed fluxes with the
model fluxes can only be made through broadband photometry. Any
attempt to determine atmospheric parameters from spectroscopic data
alone should be considered extremely dangerous and will not be
attempted in this analysis. For instance, the drop in the infrared
flux appears much more severe in the spectrum of \sdss\ than what is
actually measured in photometry. Most likely this is due to the
difficulty of determining an accurate instrumental response where
throughput is low.

The lower panel of Figure \ref{fg:f1} compares the relative energy
distributions of \sdss\ and \lhs\ normalized at the $R$ bandpass. The
resemblance of both objects is remarkable, and they will thus be
considered together in the following model atmosphere analysis.

\section{Model Atmosphere Analysis}

Our white dwarf model atmospheres are based on the calculations of 
\citet{bsw95}, with the improvements described in \citet{blr01}, 
\citet{ber01}, and references therein; the
{\it ad hoc} pseudo-continuum opacity originating from the Lyman edge,
introduced by \citet{ber01} to account for some missing ultraviolet
opacity, is omitted in the present calculations (see discussion
in \S~4). The models extend down to $\Te=1500$~K, and the atmospheric
composition may vary from pure hydrogen to pure helium, or any mixed
hydrogen and helium compositions. Also, heavier elements can be
included in the equation of state to study the indirect effects of
traces of metals on the predicted fluxes.

We also make use of the latest collision-induced opacity calculations
by Borysow and collaborators. In particular, we include the \htwo-He
calculations described in \citet{jorgensen}, as well as the recent
\htwo-\htwo\ opacities of \citet{borysow01}, which up to now had been
included in our models using the approximate formalism of
\citet{borysow97}. These improved calculations have a significant
effect on the predicted energy distributions, especially below
$\Te=4000$~K, as can be appreciated from Figure \ref{fg:f2} where we
compare theoretical fluxes of our pure hydrogen models at various
effective temperatures for the two sets of calculations \citep[see
also][]{rohrmann02}. The absorption bands near 0.8 and 1.1 \micron\ are
much more pronounced than previously estimated, and this will help
resolve the ambiguity between our hydrogen- and helium-rich solutions
presented below.

Prior to fitting each energy distribution in detail, we explore
variations of all atmospheric parameters, as we feel that such a
detailed investigation has never been carried out properly in the
literature. We then attempt to fit the energy distributions of \sdss\
and \lhs\ with those predicted from our model atmospheres using a
nonlinear least-squares method \citep[see][for details]{brl97}. Only
$T_{\rm eff}$ and the solid angle $(R/D)^{2}$, where $R$ is the radius
of the star and $D$ its distance from Earth, are considered free
parameters. The value of $\logg$ is either assumed, or for \lhs, the
distance can be obtained from the trigonometric parallax measurement
of $\pi=33.04\pm 0.50$ mas \citep{harris99}, and the resulting stellar
radius $R$ converted into mass (and thus $\logg$) using evolutionary
models. Throughout we rely on the C/O-core cooling sequences described
in \citet{blr01} with thin and thick hydrogen layers, which are based
on the calculations of
\citet{fon01}. At the low effective temperatures considered here, the
thickness of the hydrogen layer does not affect the $\logg$
determination, only the cooling age estimates.

\subsection{Pure Hydrogen Composition}

The top panel of Figure \ref{fg:f3} compares the relative energy
distributions of \sdss\ and \lhs\ normalized to unity at the $R$
bandpass with pure hydrogen models at $\logg=8.0$ normalized to unity
at the maximum flux. This comparison shows that pure hydrogen models
fail to match simultaneously the blue and infrared portions of the
observed energy distributions. While the overall shape of the infrared
flux distribution is qualitatively well reproduced with a model near
$\Te\sim 2500$~K, the blue portion is predicted much too steep at that
temperature. Similarly, the $B$ and $V$ photometry is better
reproduced with the coolest model at 1500~K, but the infrared fluxes
are then predicted too low.

The effects of varying the surface gravity are shown in the middle
panel of Figure \ref{fg:f3} for models at $\Te=2500$~K. While the
infrared fluxes are very sensitive to variations in $\logg$, in
particular near the $I$ bandpass, the monochromatic fluxes shortward
of $R$ remain unaffected. Experiments with different effective
temperatures (not shown here) indicate that it is never possible to
reproduce all broadband photometric observations simultaneously with
pure hydrogen models.

\citet{saumon99} have studied the non-ideal effects of the equation 
of state on the atmospheric structure and broadband fluxes of pure
hydrogen models \citep[see also][]{rohrmann02}. In order to study
these effects here, we have included the equation of state of
\citet{saumon95} in our model calculations. Our results (not shown 
here) compare well with those shown in Figures 1 ($\log T$ vs $\log
P$) and 3 (\mv\ vs $V-I$) of \citet{saumon99}. The effects on the
energy distribution of a $\Te=2500$~K, $\logg=8.0$ model are
illustrated in the bottom panel of Figure
\ref{fg:f3}, and they are shown to be negligible, at least in the
sense that the failure of the pure hydrogen models to reproduce the
observations in the two upper panels is not due to the neglect of the
non-ideal effects in the equation of state.

Our formal fits to the optical and infrared photometry of \sdss\ and
\lhs\ are displayed in Figure \ref{fg:f4} assuming a value of 
$\logg=8.0$ for both stars. As discussed by \citet{brl97}, when the
observed energy distribution varies considerably over the filter
bandpasses --- which is obviously the case here --- the comparison of
the model fluxes with the observed fluxes must be performed by
converting the model monochromatic fluxes into {\it average fluxes}
from an integration over the transmission function of the
corresponding bandpasses. While both of the monochromatic and average
model fluxes are shown in Figure \ref{fg:f4}, only the average fluxes
are used in the least-square fitting technique described above. The
surface gravity for \lhs\ inferred from the trigonometric parallax
measurement is well below 7.0, and our best fit at $\logg=7.0$ is
shown in Figure \ref{fg:f4} as well; the overluminous property of
\lhs\ is discussed further below. The detailed fits shown here confirm
the qualitative discussion above. {\it Our pure hydrogen models fail
to reproduce the observed energy distribution of these two very cool
white dwarfs}. In particular, the peaks of the energy distributions
are predicted much narrower than observed.

\subsection{Mixed Hydrogen and Helium Composition}

Collision induced absorptions by molecular hydrogen may also result
from collisions with neutral helium. Since cool helium-rich white
dwarfs tend to have lower opacities and thus higher atmospheric
pressures, the infrared flux deficiency appears at higher effective
temperatures in helium-rich white dwarfs than in the pure hydrogen
models considered above, as can be seen in Figure 9 of
\citet{blr01} which shows pure hydrogen and helium-rich model sequences
in the ($V$--$I$, $V$--$K$) two-color diagram.

The two upper panels of Figure \ref{fg:f5} compare the relative energy
distributions of \sdss\ and \lhs\ with models at $\Te=3250$~K,
$\logg=8.0$, for various atmospheric compositions, from pure hydrogen
to pure helium. At that temperature, the maximum infrared absorption
is reached at a hydrogen abundance of only $\nh=10^{-5}$ (shown in
both upper panels); smaller or larger hydrogen abundances yield larger
infrared fluxes. A comparison of the results shown in this figure with
those of Figure \ref{fg:f3} indicates that the helium-rich models, for
a comparable infrared depression, tend to have energy peaks that are
broader than in pure hydrogen models, in better agreement with the
observed energy distributions of both \sdss\ and \lhs. Still, it is
difficult even with helium-rich models to reconcile the optical and
infrared energy distributions of these cool white dwarfs, as confirmed
by our detailed fits discussed below.

In the bottom panel of Figure \ref{fg:f5} we also explore the
influence of traces of heavier elements on the emergent fluxes
\citep[see also][]{jorgensen,ber01}.  Model atmospheres have been
calculated by including metals in the equation-of-state only. Metal
opacities are not taken into account as the main effect of the
presence of heavy elements is to provide free electrons, and to
increase the contribution of the He$^-$ free-free opacity with respect
to Rayleigh scattering, otherwise the dominant source of opacity in
pure helium models \citep[see also][]{provencal02}. In this particular
experiment, we consider only calcium, the most common metal observed
in DZ stars. These results indicate that the shape of the energy
distribution is quite sensitive to the presence of heavier elements in
the atmosphere. If these cool white dwarfs do have extreme helium-rich
compositions, as suggested below, the presence of metals could affect
significantly the atmospheric parameter determination.

Our formal fits to the photometric observations of \sdss\ and
\lhs\ using helium-rich models are shown in Figure \ref{fg:f6},
again assuming $\logg=8.0$ for both stars.  While those fits are
somewhat better than those shown in Figure
\ref{fg:f4}, they are not perfect, and certainly not as convincing as
those shown in \citet{brl97} or \citet{blr01} for white dwarfs only a
few hundreds of degrees warmer. The peaks of the energy distributions
are still predicted too sharp, and the flux in the $R$ band is always
overestimated. Experiments with models including metals have not
improved the quality of the fits, and will not be discussed
further. Our best fit to \lhs\ using the trigonometric parallax
measurement is also displayed in Figure \ref{fg:f6}, and yields a
value of $\logg=7.27$, or a mass of only 0.23 \msun. Such a low-mass
would imply in turn that \lhs\ has a helium core --- the product of
close binary evolution. However, this interpretation is at odds with
the spectroscopic mass distributions of DA and DB stars by
\citet{bsl92} and \citet{beauchamp96}, respectively, and the
photometric analysis of cool white dwarfs with trigonometric parallax
measurements by \citet{blr01}, which indicate that low mass white
dwarfs all have hydrogen-rich atmospheres.

Our best solutions for \lhs\ with pure hydrogen or mixed
hydrogen/helium compositions are contrasted in Figure \ref{fg:f7}, and
compared with the optical spectrum. While none of our fits reproduce
the observations perfectly, the optical spectrum allows us to rule out
the pure hydrogen solution since the deep absorption feature near 0.8
\micron\ is clearly not observed. We note that calculations with the
earlier collision-induced \htwo\ opacities of \citet[][not shown
here]{borysow97} precluded us from reaching such a conclusion because
the depth of the molecular bands were not predicted as deep (see
Figure \ref{fg:f2}). We are now able to conclude with our latest model
grid that both \sdss\ and \lhs\ are better explained in terms of
extreme helium-rich compositions, despite the fact that our current
model atmospheres do not yield perfect fits to the observed
photometry.

Another argument in favor of the helium-rich solution is the
overluminosity problem for \lhs\ discussed earlier with the pure
hydrogen models.  This is best illustrated in Figure \ref{fg:f8} where
we show the location of \lhs\ in a $M_V$ vs ($V$--$I$) color-magnitude
diagram together with pure hydrogen and $\nh=10^{-5}$ models at
$\logg=8.0$. While it is clearly impossible to reconcile the observed
luminosity of \lhs\ with pure hydrogen models, the helium-rich
sequence is more consistent with the location of \lhs\ in this
diagram. Lower gravity (larger radii and thus larger luminosity)
models would improve the agreement even further, as indicated in
Figure \ref{fg:f8} where we show the predicted values of $M_V$
and ($V$--$I$) obtained from our best solution at $\logg=7.27$ shown
in the lower panel of Figure \ref{fg:f6}.

The overluminosity of \lhs\ may also suggest that these stars are
unresolved degenerate binaries. We have also attempted to fit the
observed energy distributions with two white dwarf models, weighted by
their respective radius. Despite the large number of free parameters
in the possible solutions, the best fits achieved are not better than
those shown in Figures \ref{fg:f4} or \ref{fg:f6} and thus were not
considered further.

Finally we show in Figure \ref{fg:f9} the location of \sdss\ and \lhs\
in a ($V$--$I$, $V$--$H$) two-color diagram, together with theoretical
colors from model atmospheres with various chemical compositions. Also
shown are the photometric data from the cool white dwarf sample of
\citet{brl97,blr01}, as well as the photometry of WD 0346$+$246, the 
low luminosity white dwarf discovered by \citet{hambly97}. The
variation of helium abundances is illustrated by the sequence at
$\Te=3250$~K, which performs an excursion from the pure hydrogen to a
pure helium composition, reaching minimum values of $V$--$I$ and
$V$--$H$ at $\nh\sim10^{-5}$. In this diagram, \sdss\ is consistent
with the helium-rich solution while \lhs\ is marginally consistent
with both solutions. The location of the cool white dwarf WD
0346$+$246 also suggests a helium-rich composition (see discussion
below).

\section{Discussion}

The extreme helium-rich composition inferred in our analysis for both
\sdss\ and \lhs\ poses an obvious and challenging problem.
Because the outer layers of cool white dwarfs are strongly convective,
the hydrogen and helium chemical composition tends to be more or less
homogeneous throughout the mixed convection zone. Since below $\Te\sim
12,000$~K, the mass of the deep helium convection zone is almost
constant at $M_{\rm He-conv}\sim 10^{-6}~M_{\ast}$ \citep{tassoul90},
the small hydrogen abundances of only $\nh\sim10^{-5}-10^{-4}$ derived
in our analysis imply a {\it total} hydrogen mass of
$\sim5\times10^{-12}$ \msun. If this amount of hydrogen has been
accreted from the interstellar medium over a cooling age of roughly 10
Gyr for a 3500~K white dwarf, the implied accretion rate would be only
$\sim 10^{-22}$ \msun\ yr$^{-1}$, a value that is completely
unrealistic. For instance, the theoretical estimates of \citet{wes79}
suggest time-averaged accretion rates of $\sim10^{-17}$ \msun\
yr$^{-1}$. Of course, the extreme helium-rich compositions derived
here for \sdss\ and \lhs\ cannot be completely ruled out on the basis
of these arguments alone since the problem of the accretion of
hydrogen in cool white dwarfs is a long standing one, and no
satisfactory explanation has yet been proposed to account for the
persistence of helium-rich white dwarfs at low effective temperatures.

We note in this context that WD 0346+246 shown in Figure \ref{fg:f9}
has been interpreted by \citet{opp01b} as a $\Te=3750$~K white dwarf
with an extremely small hydrogen abundance of $\log \nh=-6.4$. A
reanalysis of this object by \citet{ber01} with the improved \htwo-He
collision-induced opacities of \citet{jorgensen} indicates an even
lower hydrogen abundance of $\nh\sim10^{-9}$. Because this solution
was improbable --- but not impossible --- from accretion
considerations, \citet{ber01} proposed an alternative solution with a
higher hydrogen abundance of $\nh\sim0.8$, but was also forced to
introduce in the model calculations a bound-free opacity from the
Lyman edge associated with the so-called dissolved atomic levels of
the hydrogen atom, in order to reduce the near-UV flux and match the
optical photometric observations.  This additional pseudo-continuum
opacity has not been included in the present calculations as the UV
flux is already predicted too low for \sdss\ and \lhs.

While it is probably safe to conclude that these cool white dwarfs have
helium-rich atmospheres and effective temperatures below 4000~K, it is
not yet possible to determine their atmospheric parameters and ages
with great precision since the models fail to reproduce the
photometric observations in detail.  The large parameter space
explored in our analysis suggests that the source of this discrepancy
lies in the physics included in our model atmospheres, which is either
inadequate or incomplete. One avenue of investigation worth
considering is non-ideal effects of the equation-of-state at the high
atmospheric pressures that characterize helium-rich atmospheres. So
far, these effects have only been estimated in pure hydrogen or pure
helium atmospheres
\citep[][this paper]{bsw95,saumon99,rohrmann02}. It has also been
demonstrated recently by \citet{iglesias02} that the helium free-free
opacity and Rayleigh scattering at high densities typical of those
encountered here are seriously overestimated. The use of their
improved treatment in model atmosphere calculations would most likely
increase the flux where the collision-induced opacity is less
important, i.e.~for wavelengths below 0.7 \micron, precisely in the
region where we observe the largest discrepancy. This could even solve
the overluminosity problem of \lhs\ altogether as the calculated
absolute visual magnitude for helium-rich white dwarfs would be
brighter.

All white dwarfs below $\Te=4000$~K for which a detailed analysis has
been performed have been explained in terms of some mixed hydrogen
and helium atmospheric compositions, or even a pure helium composition
\citep[][]{harris99,opp01b,ber01,farihi02}. There are still
no white dwarfs below $\Te=4000$~K that have been successfully and
convincingly explained in terms of a pure hydrogen atmospheric
composition. This result may not be completely unexpected since
helium-atmosphere white dwarfs with their lower opacities have cooling
time scales that can be considerably shorter than their
hydrogen-atmosphere counterparts
\citep[see, e.g.,][]{hansen99, blr01}. Hence 
for a given population and at a given mass, the coolest white dwarfs
are expected to have helium-dominated atmospheres. For instance, a
cool helium-rich white dwarf at $\Te=3250$~K (i.e., our temperature
estimate for \sdss) with an average mass of $\sim 0.7$ \msun\ 
\citep[see Fig.~22 of][]{blr01} has a
cooling age of only 9.4 Gyr according to our evolutionary models with
thin hydrogen layers \citep[see also Fig.~24 of][]{blr01}; by
comparison, a similar object with a thick hydrogen layer would have an
age of 11.6 Gyr. This 9.4 Gyr estimate is entirely consistent with the
age of the oldest objects in the Galactic disk analyzed by
\citet{blr01}, and the so-called ultracool white dwarfs reported in
the literature may not be terribly old after all. Although WD
0346$+$246 has kinematics supporting halo membership
\citep{hambly99}, \lhs\ has a lower tangential velocity consistent
with membership of either disk or halo, and \sdss\ has an even lower
velocity inconsistent with halo membership \citep{harris01}.

Despite the fact that we do not fully understand the accretion
mechanism in cool white dwarfs, helium-atmosphere white dwarfs must
eventually accrete ``some amount'' of hydrogen, at which point they
will exhibit a strong infrared flux deficiency, even if the amount of
hydrogen is extremely small according to our calculations. \sdss\ and
\lhs\ are most likely such objects, and extremely cool
hydrogen-atmosphere white dwarfs belonging to the old halo population
of the Galaxy are yet to be identified.

\acknowledgements We are grateful to H.~Harris for providing us with 
the spectra of \sdss\ and \lhs. This work was supported in part by the
NSERC Canada and by the Fund NATEQ (Qu\'ebec). The $J$ and $H$
magnitudes for \sdss\ were obtained through the UKIRT Service
Programme. UKIRT, the United Kingdom Infrared Telescope, is operated
by the Joint Astronomy Centre on behalf of the U.K. Particle Physics
and Astronomy Research Council.

\clearpage

\figcaption[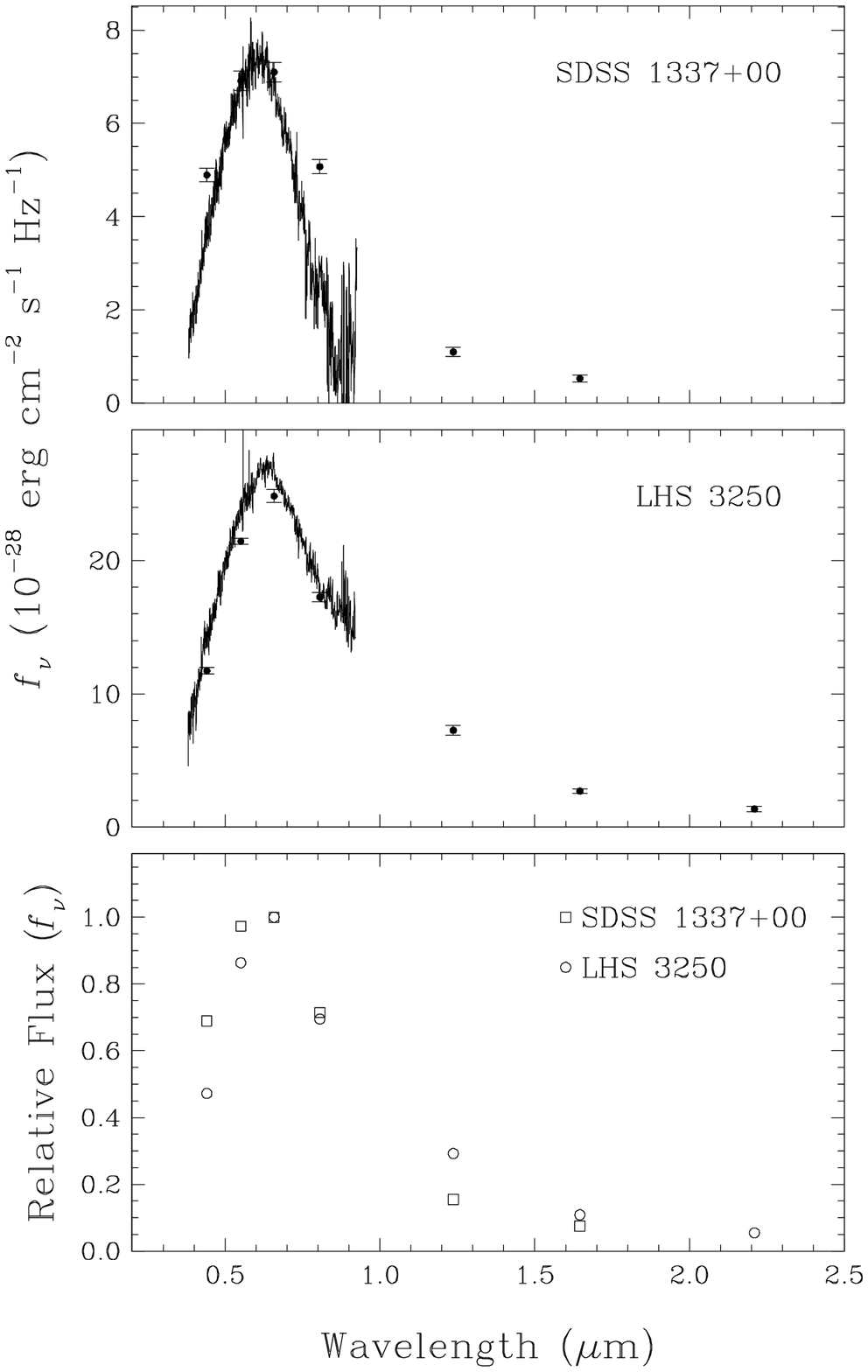] {Upper panels: Optical $BVRI$ and infrared $JHK$
photometry of \sdss\ and \lhs\ displayed as broadband fluxes (with
corresponding error bars) following the prescription of
\citet[][]{brl97}; only $J$ and $H$ have been measured for \sdss. Also
shown for comparison are the spectroscopic observations for each
object taken from \citet{harris01}. Bottom panel: Comparison of the
energy distributions of both stars normalized to unity at
$R$.\label{fg:f1}}

\figcaption[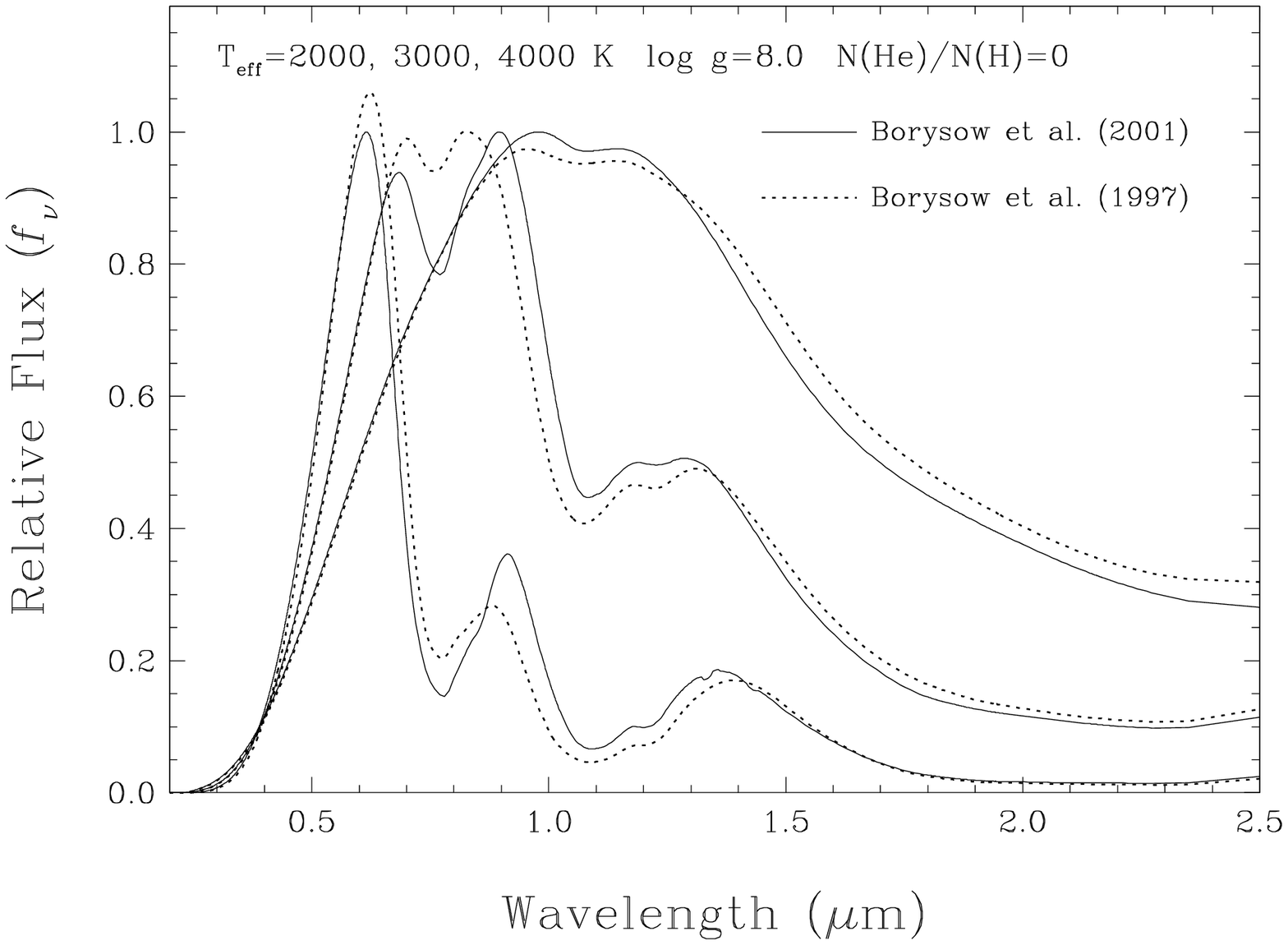] {Comparison of emergent fluxes from our pure hydrogen 
model atmospheres using the collision-induced opacity calculations of
\citet[][dotted lines]{borysow97} and 
\citet[][solid lines]{borysow01}. For each effective temperature 
(the leftmost model is at $\Te=2000$~K), the fluxes are normalized to
unity at the maximum flux of the \citet{borysow01}
models.\label{fg:f2}}

\figcaption[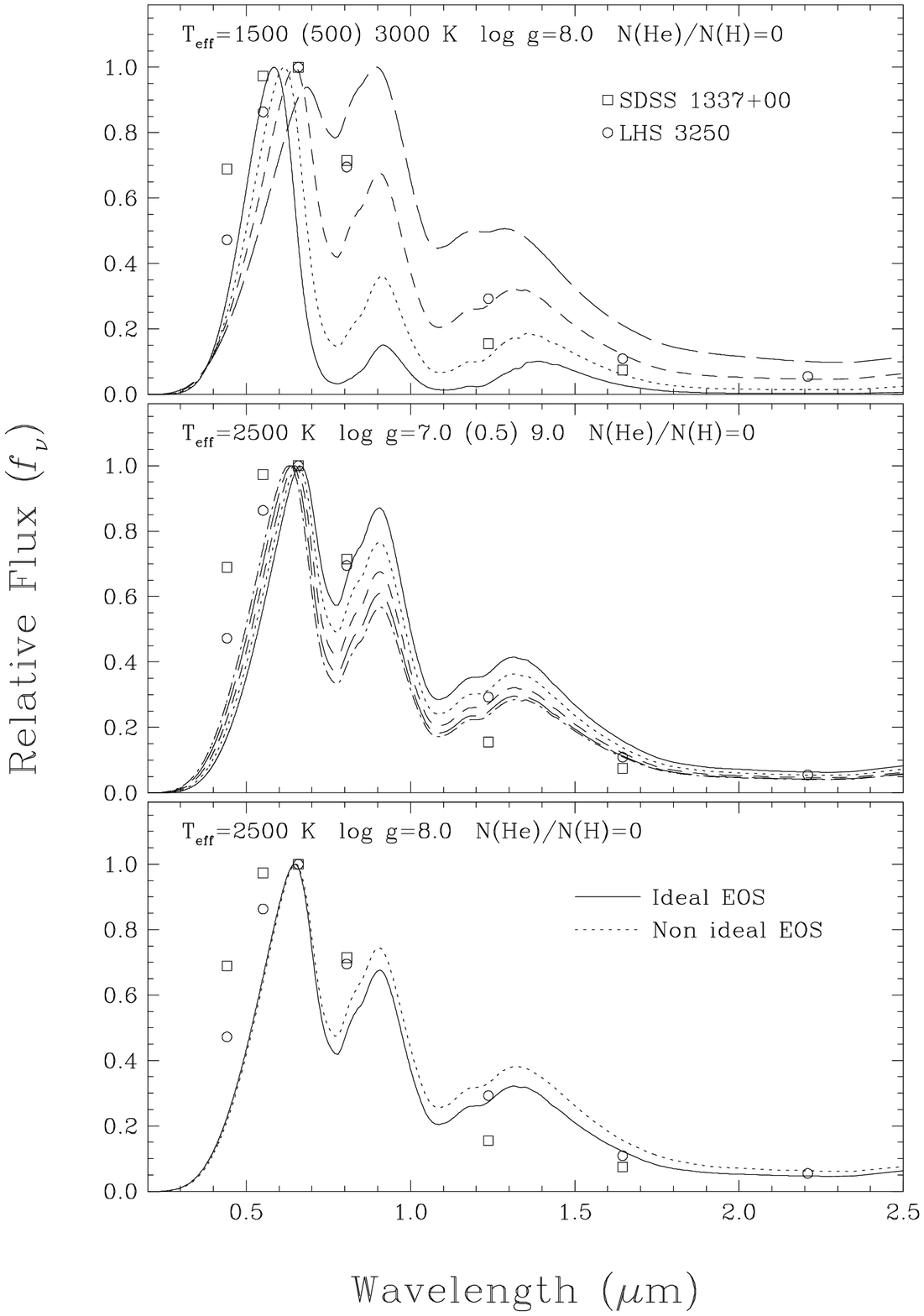] {Relative energy
distributions of \sdss\ and \lhs\ normalized to unity at the $R$ bandpass
compared with various pure hydrogen models normalized to unity at the
maximum flux.  The top panel compares models at $\logg=8.0$ with
$\Te=1500$~K (solid line) to 3000 K by step of 500 K. The middle panel
compares models at $\Te=2500$~K with $\logg=7.0$ (solid line) to 9.0
by step of 0.5. The bottom panel compares models at $\Te=2500$~K,
$\logg=8.0$ calculated with an ideal equation of state, as well as
with the non-ideal equation of state of
\citet{saumon95}.\label{fg:f3}}

\figcaption[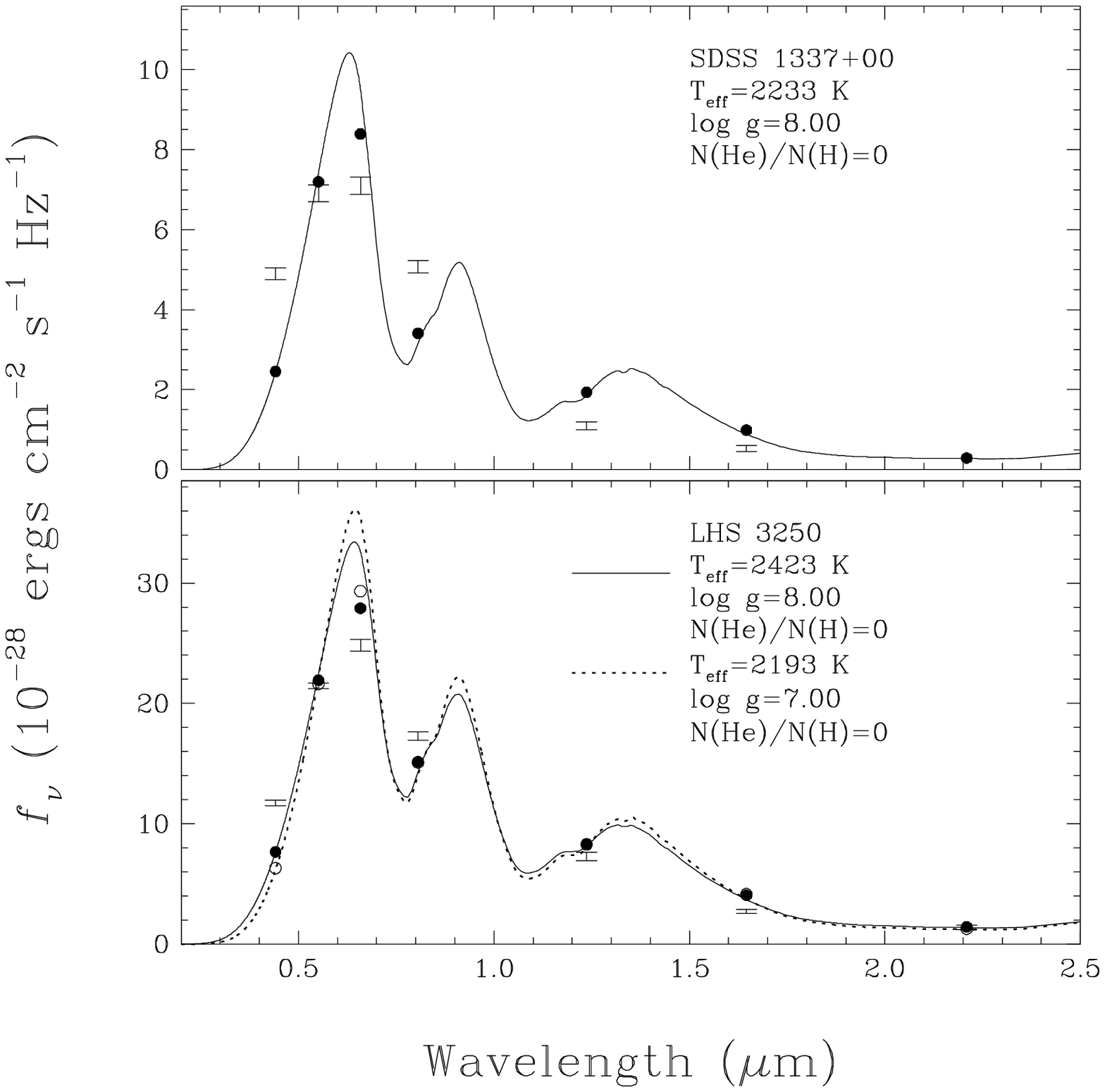] {Fits to the energy distributions of \sdss\ and 
\lhs\ with pure hydrogen models. The optical $BVRI$ 
and infrared $JH(K)$ photometric observations are shown by the error
bars. The solid lines correspond to the model monochromatic fluxes at
$\logg=8.0$, while the filled circles represent the average over the
filter bandpasses. For \lhs, our best fit at $\logg=7.0$ is shown as
the dotted line and open circles.\label{fg:f4}}

\figcaption[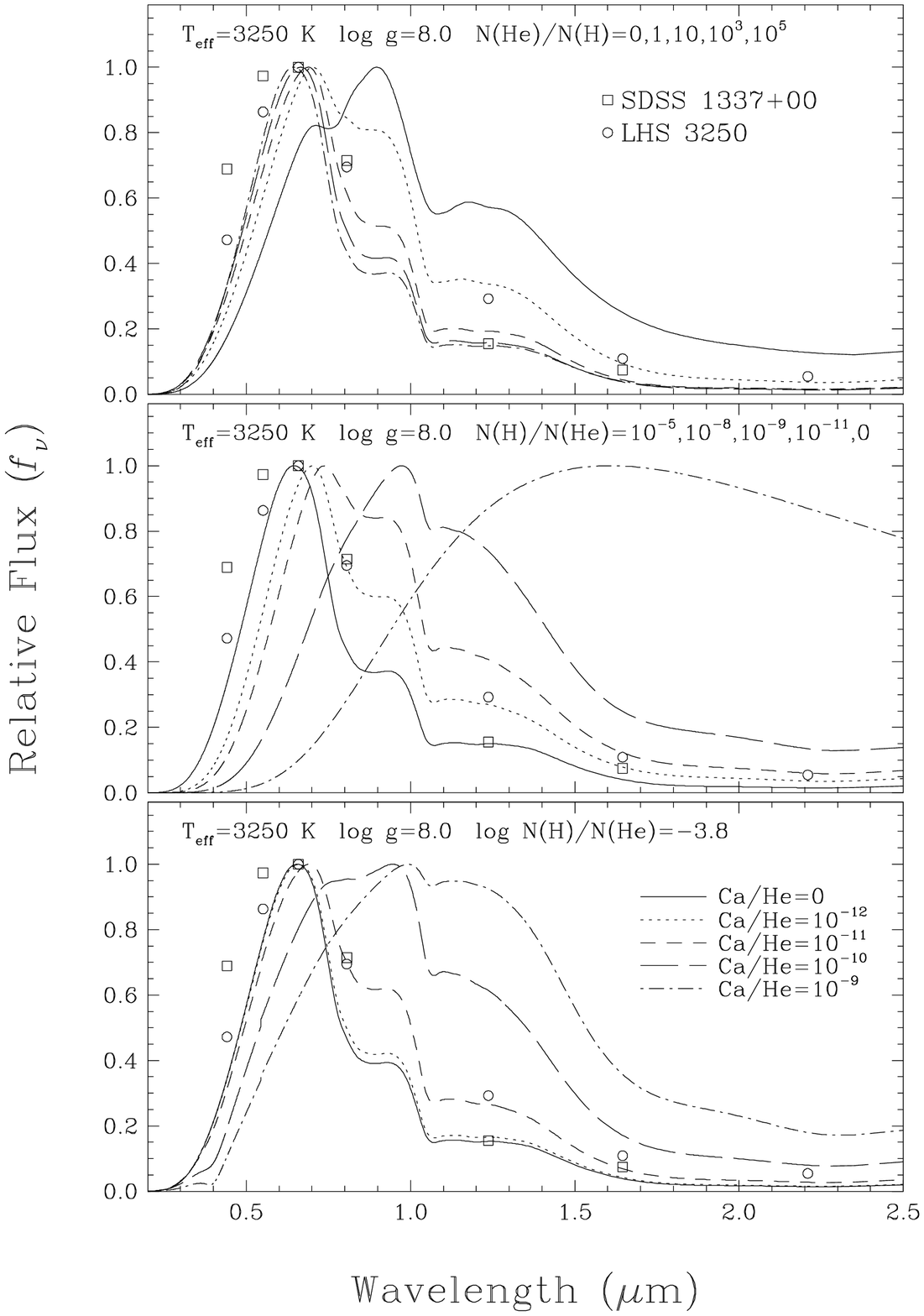] {Relative energy
distributions of \sdss\ and \lhs\ normalized to unity at the $R$ bandpass
compared with various mixed hydrogen/helium/calcium models normalized
to unity at the maximum flux. The top panel compares models at
$\Te=3250$~K and $\logg=8.0$ from a pure hydrogen composition (solid
line) to a value of $\nhe=10^5$ where the infrared flux deficiency is
the strongest. In the middle panel, the hydrogen abundance is further
decreased from a value of $\nh=10^{-5}$ (solid line) to a pure helium
composition. In the bottom panel, the helium abundance is fixed but
the calcium abundance is varied.\label{fg:f5}}

\figcaption[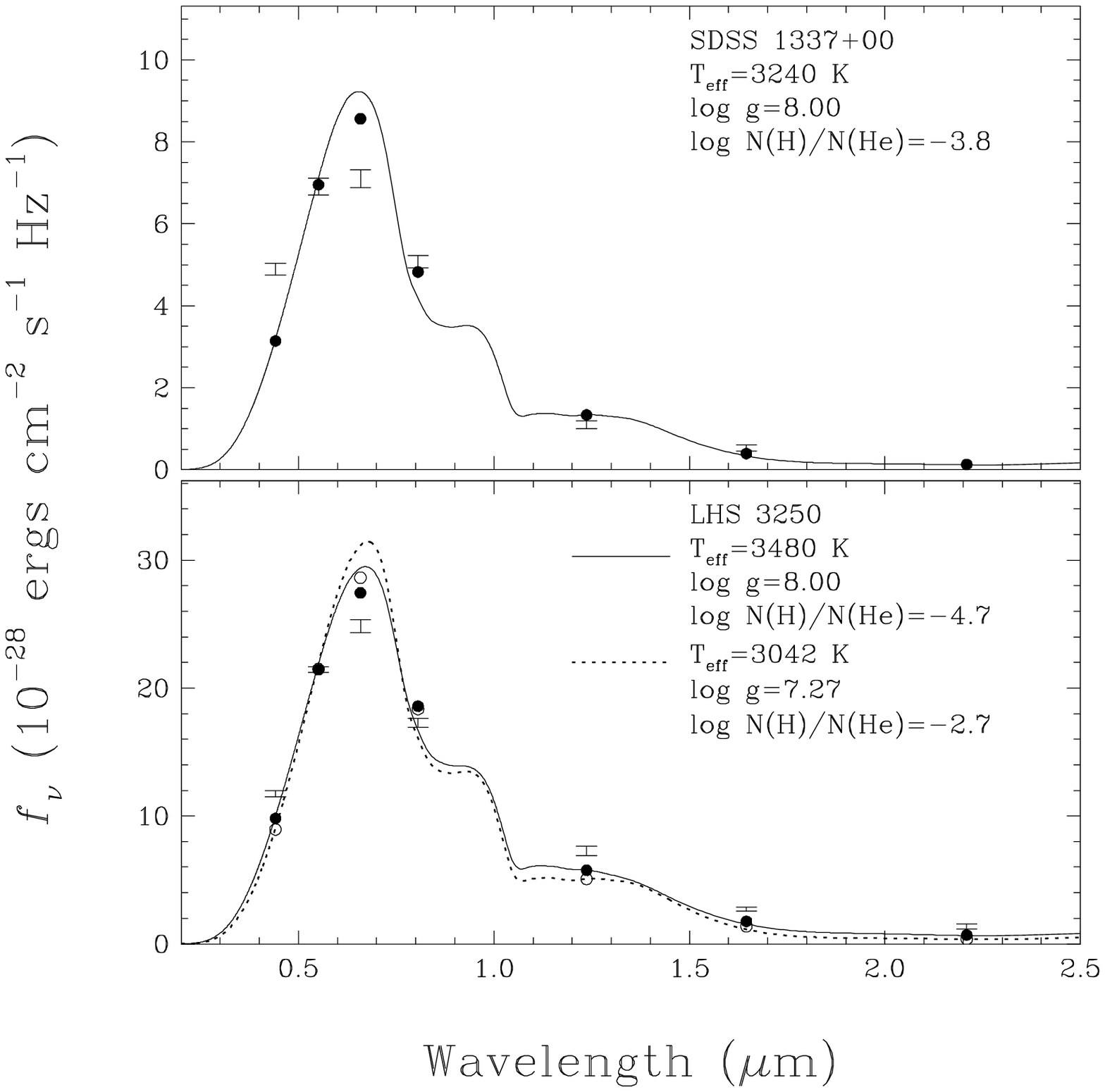] {Fits to the energy distributions of \sdss\ and 
\lhs\ with pure helium-rich models. The various symbols are explained 
in Figure \ref{fg:f4}. The fit for \lhs\ shown as the dotted line and
open circles relies on the trigonometric parallax
measurement.\label{fg:f6}}

\figcaption[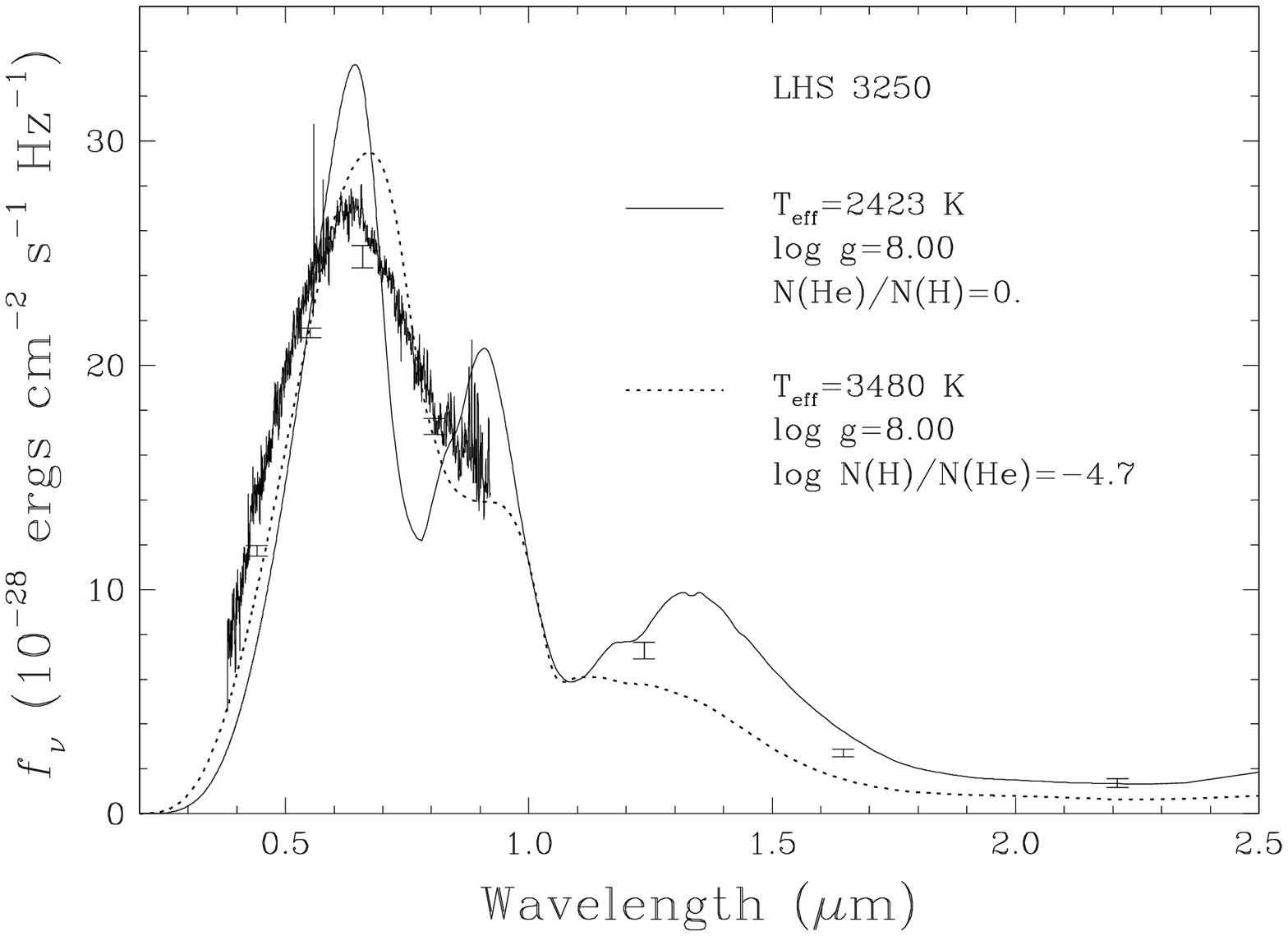] {Comparison of the best solutions for \lhs\ under the 
assumption of a pure hydrogen composition ({\it solid line}) and a
mixed hydrogen/helium composition ({\it dotted line}). Also shown are
the broadband photometry and optical spectrum. The latter suggests
that \lhs\ has a helium-rich composition rather than a pure hydrogen
atmosphere.\label{fg:f7}}

\figcaption[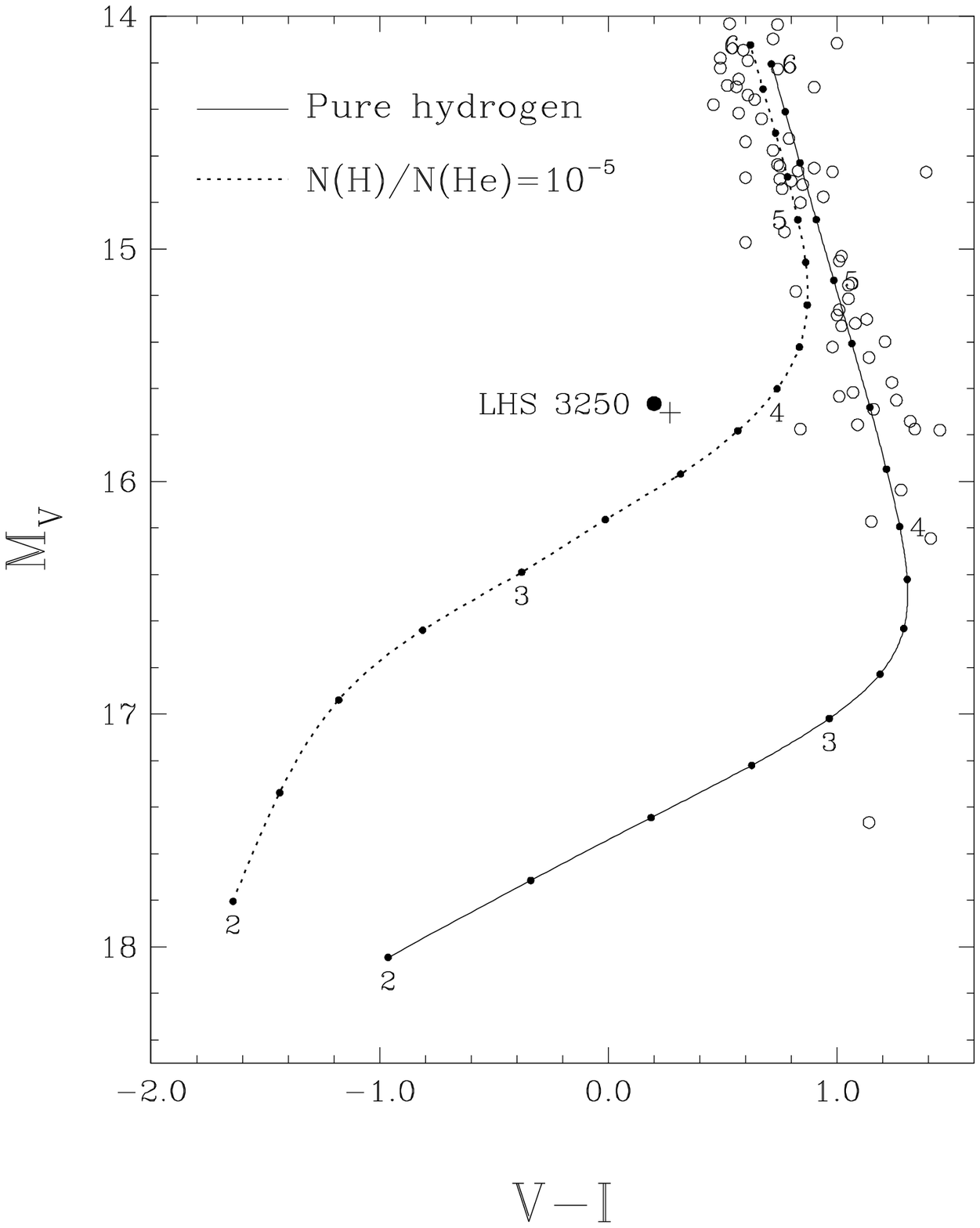] {Location of \lhs\ in the \mv\ vs ($V$--$I$) 
color-magnitude diagram; the size of the solid dot corresponds to the
error of \mv. The open circles represent the photometric data taken from
\citet{blr01}. The model sequences correspond to $\logg=8.0$ 
models for pure hydrogen and $\nh=10^{-5}$ atmospheric
compositions. The small dots on each sequence are separated by
$\Delta\Te=250$~K, and effective temperatures are indicated in units
of $10^3$~K. The plus sign indicates the location of our best solution
at $\logg=7.27$ shown in the lower panel of Figure \ref{fg:f6}.
\label{fg:f8}}

\figcaption[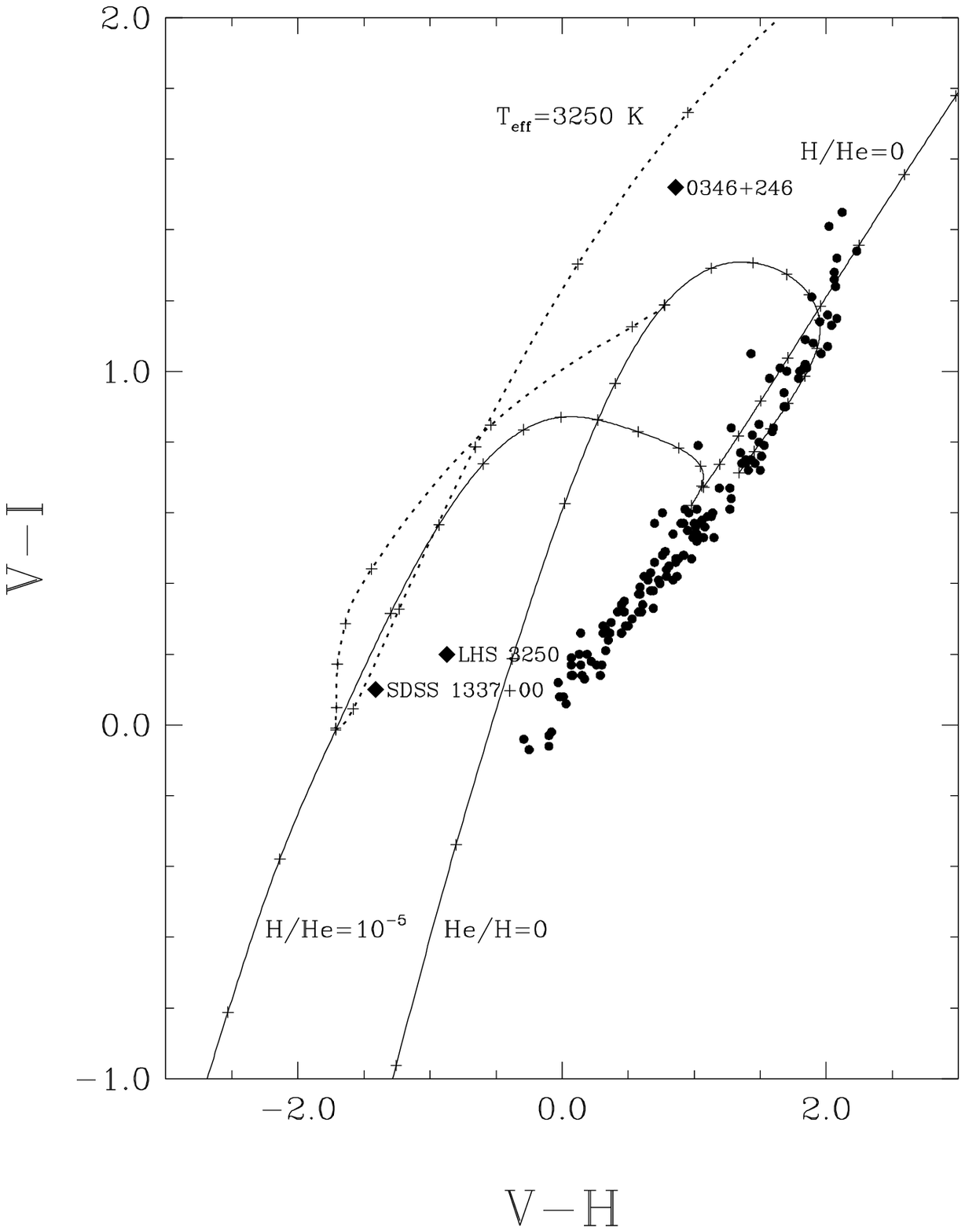] {Location of \sdss\ and \lhs\ in the
($V$--$I$, $V$--$H$) two-color diagram. Also shown are the cool white
dwarf photometric observations taken from \citet{brl97,blr01}, and the
photometry of WD 0346$+$246 from \citet{opp01b}. The theoretical
colors shown as solid lines correspond to pure hydrogen models (${\rm
He}/{\rm H}=0$), pure helium models (${\rm H}/{\rm He}=0$), and mixed
hydrogen and helium compositions (${\rm H}/{\rm He}=10^{-5}$) at
various effective temperatures, all at $\logg=8.0$; on each sequence
the plus signs are separated by $\Delta\Te=250$~K, starting at
$\Te=6000$~K on the observed photometric sequence. The dotted line
corresponds to a $\Te=3250$~K, $\logg=8.0$ model sequence for various
helium abundances; the plus signs are separated by 1 dex in $\log
\nhe$, starting at $\nhe=10^{-2}$ near the pure hydrogen sequence.
\label{fg:f9}}

\clearpage
\begin{figure}[p]
\plotone{f1.ps}
\begin{flushright}
Figure \ref{fg:f1}
\end{flushright}
\end{figure}

\clearpage
\begin{figure}[p]
\plotone{f2.ps}
\begin{flushright}
Figure \ref{fg:f2}
\end{flushright}
\end{figure}

\clearpage
\begin{figure}[p]
\plotone{f3.ps}
\begin{flushright}
Figure \ref{fg:f3}
\end{flushright}
\end{figure}

\clearpage
\begin{figure}[p]
\plotone{f4.ps}
\begin{flushright}
Figure \ref{fg:f4}
\end{flushright}
\end{figure}

\clearpage
\begin{figure}[p]
\plotone{f5.ps}
\begin{flushright}
Figure \ref{fg:f5}
\end{flushright}
\end{figure}

\clearpage
\begin{figure}[p]
\plotone{f6.ps}
\begin{flushright}
Figure \ref{fg:f6}
\end{flushright}
\end{figure}

\clearpage
\begin{figure}[p]
\plotone{f7.ps}
\begin{flushright}
Figure \ref{fg:f7}
\end{flushright}
\end{figure}

\clearpage
\begin{figure}[p]
\plotone{f8.ps}
\begin{flushright}
Figure \ref{fg:f8}
\end{flushright}
\end{figure}

\clearpage
\begin{figure}[p]
\plotone{f9.ps}
\begin{flushright}
Figure \ref{fg:f9}
\end{flushright}
\end{figure}

\end{document}